\documentclass[aps,pra,twocolumn,nofootinbib,longbibliography,amssymb]{revtex4-1}
\usepackage{amsmath}
\usepackage{braket}
\usepackage{graphicx}
\usepackage[shortlabels]{enumitem}
\usepackage{substr}
\usepackage{verbatim}
\allowdisplaybreaks    
\begin{document}

\title{On the validity of many-mode Floquet theory with commensurate frequencies}
\author{A.~N.~Poertner}
\author{J.~D.~D.~Martin}
\affiliation{Department of Physics and Astronomy, University of Waterloo, Waterloo, Canada}

\date{\today}

\begin{abstract}

Many-mode Floquet theory [T.-S. Ho, S.-I. Chu, and J. V. Tietz, Chem. Phys. Lett. \textbf{96}, 464 (1983)] is a technique for solving the time-dependent Schr\"odinger equation in the special case of multiple periodic fields, but its limitations are not well understood.  We show that for a Hamiltonian consisting of two time-periodic couplings of commensurate frequencies (integer multiples of a common frequency), many-mode Floquet theory provides a correct expression for unitary time evolution.  However, caution must be taken in the interpretation of the eigenvalues and eigenvectors of the corresponding many-mode Floquet Hamiltonian, as only part of its spectrum is directly relevant to time evolution.  We give a physical interpretation for the remainder of the spectrum of the Hamiltonian.  These results are relevant to the engineering of quantum systems using multiple controllable periodic fields.

\end{abstract}


\maketitle

\section{Introduction}  \label{se:jddm_introduction}

In quantum mechanics there is hardly a task more fundamental than solving the time-dependent Schr\"{o}dinger equation.  A particularly important case is atomic evolution in the presence of classically prescribed electromagnetic fields, corresponding to Hamiltonians of the form: $H(t) = H_0 + V(t)$, where the time-independent $H_0$ describes the atomic system in the absence of the fields, and a (possibly) time-varying $V(t)$ accounts for the presence of the fields.

The frequent situation that the fields are periodic in time may be deftly handled using Floquet theory:  suppose that there is a single relevant time-dependent field, periodic in time, so that $V(t) = V(t+T)$ for some period $T$ and for all times $t$.   If a finite basis of dimension $N_A$ may be used to describe the atomic system, Floquet theory tells us that there are $N_A$ independent solutions for the state vector of the form \cite{shortdoi:czxdv8}:
\begin{equation} \label{eq:each_floquet_soln}
\ket{\psi_j(t)} = e^{-iE_j t / \hbar} \ket{\phi_j(t)},
\end{equation}
where we have labelled each of the solutions with index $j$. The $E_j$'s are known as the {\em quasi-energies} and the corresponding $\ket{\phi_j(t)}$'s --- so-called {\em quasi-states} --- have the same periodicity as the Hamiltonian: $\ket{\phi_j(t)} = \ket{\phi_j(t + T)}$.  This periodicity suggests a Fourier expansion:
\begin{equation} \label{eq:quasi_fourier_expansion}
\ket{\phi_j(t)}=\sum_n \ket{\tilde{\phi}_j(n)} e^{i n \omega t}
\end{equation}
where $\omega=2\pi/T$.  Shirley  \cite{shortdoi:czxdv8} showed that when the time-dependent Schr\"{o}dinger equation (TDSE) is expressed in terms of the $\ket{\tilde{\phi}_j(n)}$ expansion ``coefficients'', all $N_A$ of the solutions --- in form of Eq.~\ref{eq:each_floquet_soln} --- may be determined from the eigenvalues and eigenvectors of a {\em time-independent} matrix (the ``Floquet'' Hamiltonian).  Once all of the solutions are known, it is straightforward to write the unitary time evolution operator, constituting a complete solution for the quantum mechanical evolution of the atomic system in the presence of the periodic field.

In addition to having a certain aesthetic appeal, Shirley's formulation of Floquet theory (SFT) is often well-suited for explicit computations, as it may just involve a straightforward generalization of a simpler time-independent problem (for an example in Rydberg atom physics see Ref.~\cite{shortdoi:b29hxs}).

Here we are concerned with a generalization of SFT to two (or more) fields of {\em different} periodicities; for example, $H(t) = H_0 + V_1(t) + V_2(t)$ where $V_1(t) = V_1(t+T_1)$ and $V_2(t) = V_2(t+T_2)$ for all $t$, and $T_1 \ne T_2$.  If the ratio of the corresponding frequencies $f_1=1/T_1$ and $f_2=1/T_2$ may be represented as: $f_1/f_2 = N_1/N_2$ where $N_1$ and $N_2$ are integers --- so-called {\em commensurate} frequencies --- a period common to both $V_1(t)$ and $V_2(t)$ exists ($T=N_1/f_1=N_2/f_2$). Thus this situation is completely handled by SFT, albeit awkwardly --- the couplings due to each of the fields are at (different) harmonics of the common base frequency $1/T$, the details depending on $N_1$ and $N_2$.

As an alternative, Ho {\it et al.}~\cite{shortdoi:fg723q} extended SFT in a way that removes explicit references to $N_1$ and $N_2$, thereby recovering the elegance and simplicity of SFT for a field of a single periodicity.  In a similar manner to SFT, this formulation involves a unitary time evolution operator written in terms of a {\em time-independent} many-mode Floquet theory (MMFT) Hamiltonian.

The MMFT formulation has been used for nuclear magnetic resonance \cite{shortdoi:dw9t7w}, dressed potentials for cold atoms \cite{shortdoi:gcx7j9}, microwave dressing of Rydberg atoms \cite{shortdoi:gc82gc}, and superconducting qubits \cite{shortdoi:gfth9n}, to name but a few examples.  Nonetheless, independent groups have questioned the validity of MMFT \cite{shortdoi:dcs7gd,shortdoi:cmtcwh} and the completeness \cite{shortdoi:cfhh} of the justification of MMFT given in Ref.~\cite{shortdoi:fg723q}.  Subsequent publications \cite{shortdoi:b6r5dc, shortdoi:ffc2nc} by one of the authors of the original MMFT paper \cite{shortdoi:fg723q} support the conjecture \cite{shortdoi:dcs7gd} that the MMFT formulation is approximately correct in some commensurate cases, but is entirely correct for incommensurate cases (irrational frequency ratios), in dissonance with the justification presented in Ref.~\cite{shortdoi:fg723q} which is based on commensurate frequencies.

Prompted by the recent use of MMFT in a Rydberg atom study \cite{shortdoi:gc82gc}, we began to consider its correctness, particularly for two commensurate frequencies described by low $N_1$ and $N_2$, which are often relatively easy to simultaneously generate in an experiment (i.e.~as low harmonics of a common frequency source).  We computed the time evolution of a simple system in the case of commensurate frequencies numerically using MMFT and compared our results to both SFT and direct integration of the TDSE and were surprised to find no differences (when adequate basis sizes, time steps, etc.~were chosen).  This agreement is at apparent odds with the literature questioning the general applicability of MMFT and our own expectations after examination of the justification of MMFT given in Ref.~\cite{shortdoi:fg723q}. We found this situation confusing, to say the least.

In this work, we resolve these discrepancies by showing that MMFT may be used to correctly compute time evolution, {\em and that this is consistent with} the fact that not all of the eigenpairs\footnote{We refer to an eigenvector and its associated eigenvalue collectively as an {\em eigenpair}.} of the MMFT Hamiltonian correspond to the Floquet quasi-energies and quasi-states (i.e.~the $E_j$'s and $\ket{\phi_j(t)}$'s of Eq.~\ref{eq:each_floquet_soln}).

The case of incommensurate frequencies (see, for example, Ref.'s \cite{shortdoi:gcgdhz} and \cite{shortdoi:gfwhfg}) is beyond the scope of this work.

Many readers will be familiar with the background on Shirley's formulation of Floquet theory (SFT) \cite{shortdoi:czxdv8} that we review in Section \ref{se:floquet}, but perhaps less so with Ho {\it et al.'s} \cite{shortdoi:fg723q} MMFT theory, as reviewed in Section \ref{se:mmft}.  We include these sections for completeness and to establish notation.  Our results are in Section \ref{se:equivalence}, where we show how the SFT and MMFT approaches may be considered to be equivalent, and address the concerns with MMFT raised in the literature \cite{shortdoi:dcs7gd,shortdoi:cmtcwh, shortdoi:cfhh}. Section \ref{se:discussion} concludes with a summary and a discussion of the utility of MMFT in the case of commensurate frequencies.

\section{Background} \label{se:background}

\subsection{Floquet theory}  \label{se:floquet}

As a foundation for discussion of the multiple-frequency case, this section reviews Floquet theory as it applies to the solution of the TDSE:
\begin{equation} \label{eq:tdse}
  i\hbar \frac{d}{dt} \ket{\psi(t)} = \hat{H}(t) \ket{\psi(t)}
\end{equation}
given a Hamiltonian that is both periodic $\hat{H}(t) = \hat{H}(t+T)$ and Hermitian $\hat{H}^{\dagger}(t) = \hat{H}(t)$ for all times $t$.  To simplify --- but not restrict the results in a fundamental way --- the state vectors $\ket{\psi(t)}$ will be considered as belonging to a finite-dimensional inner-product space $A$ of dimension $N_A$.  In what follows this shall be referred to as the {\em atomic space}.

Since the Hamiltonian is Hermitian, we may define a unitary time evolution operator $\hat{U}(t_2, t_1)$ satisfying
\begin{equation}
  \ket{\psi(t_2)} = \hat{U}(t_2,t_1) \ket{\psi(t_1)}
\end{equation}
for all $t_1$ and $t_2$.

Floquet theory is slightly more general than required here --- the general theory is not restricted to unitary time evolution (see, for example, Ref.~\cite{isbn:978-0-471-86003-7}).  For the unitary case, Floquet theory implies
(see, for example, Ref.~\cite{shortdoi:gfw8sp} or Supplemental Materials \cite{supplemental}) that the quasi-states of Eq.~\ref{eq:each_floquet_soln} exist and may be combined to give:
\begin{equation} \label{eq:floquet_operator}
\hat{U}(t, 0) = \sum_{j=1}^{N_A} \ket{\phi_j(t)} e^{-i E_j t / \hbar} \bra{\phi_j(0)}.
\end{equation}
The quasi-energies and corresponding quasi-states may be determined by direct numerical integration of the TDSE over the duration of a single period $T$ (see, for example, Ref.~\cite{shortdoi:gcx7kb}).  However, there are alternatives to direct integration, namely SFT \cite{shortdoi:czxdv8} and MMFT \cite{shortdoi:fg723q}, which we shall now review.

\subsection{Shirley's formulation of Floquet theory (SFT)}  \label{se:shirley}

\subsubsection{The SFT Hamiltonian}

The use of Fourier decomposition to find Floquet-type solutions (e.g.~Eq.~\ref{eq:quasi_fourier_expansion}) has a long history, originating with Hill's theory regarding the motion of the moon (see, for example, Ref.~\cite{shortdoi:bbf6br}).  Following earlier more specific work by Autler and Townes \cite{shortdoi:ctpnmz}, Shirley \cite{shortdoi:czxdv8} applied these ideas to the unitary time evolution of quantum mechanics, showing that determination of the quasi-energies and quasi-states reduces to a linear eigenvalue problem similar to the normal eigenvalue problem $\hat{H} \ket{\psi} = E \ket{\psi}$ for {\em time-independent} Hamiltonians. In this section, we reproduce Shirley's Floquet theory (SFT) using a slightly modified notation suitable for extension to MMFT (similar in spirit to that of Ref.~\cite{shortdoi:b7t3tw}).

Consider an infinite-dimensional inner-product space $F$ for Fourier decomposition, spanned by an orthonormal basis set: $\{ \ket{n}_F \mid n \in \mathbb{Z} \}$, where $\mathbb{Z}$ refers to the set of all integers and $\braket{m|n}_F = \delta_{m,n}$.  The full time-dependence of the quasi-states of Eq.~\ref{eq:quasi_fourier_expansion} will be represented using a time-dependent superposition of {\em time-independent} vectors from the tensor product space $F \otimes A$:
\begin{equation} \label{eq:general_fa_solution}
\ket{\phi_j(t)}_A =
 \sum_{n = -\infty}^{\infty} e^{i n \omega t} \:\:
   \left[ \bra{n}_F \otimes \hat{I}_A
   \: \right] \:\:
   \ket{\phi_j}_{F \otimes A}
\end{equation}
where $\hat{I}_A$ is the identity in the atomic space $A$.  (Hitherto all operators and vectors were in the atomic space; henceforth we will be explicit, and for clarity avoid referring to vectors in $F \otimes A$ as ``states''.)

Vectors in $F \otimes A$ may be decomposed using the basis sets for $F$ and $A$:
\begin{equation} \label{eq:quasi_components}
\ket{\phi_j}_{F \otimes A} = \sum_{m,\alpha}
                    D_j(m, \alpha)\: \ket{m}_F \otimes \ket{\alpha}_A,
\end{equation}
where the expansion coefficients $D_j(m,\alpha)$ are complex numbers, and here and after summations over Fourier indices are implicitly from $-\infty$ to $\infty$.

We may determine the quasi-energies and expansion coefficients for a specific problem by substitution of
\begin{equation} \label{eq:time_evolving_qs}
\ket{\psi_j(t)}_A = e^{-i E_j t}
 \sum_{n} e^{i n \omega t} \:\:
   \left[ \bra{n}_F \otimes \hat{I}_A
   \: \right] \:\:
   \ket{\phi_j}_{F \otimes A}
\end{equation}
into the TDSE (Eq.~\ref{eq:tdse} with $\hbar=1$ and hereafter) and Fourier expanding the Hamiltonian: $\hat{H}_A(t) = \sum_{m} \tilde{H}_A(m) e^{i m \omega t}$.  The result \cite{shortdoi:czxdv8} is a linear eigenvalue problem (see Supplemental Materials \cite{supplemental}):
\begin{equation}
\hat{H}_{F \otimes A} \ket{\phi_j}_{F \otimes A} = E_j \ket{\phi_j}_{F \otimes A}
\end{equation}
where the {\em Floquet Hamiltonian} $\hat{H}_{F \otimes A}$ is:
\begin{align} \label{eq:floquet_hamiltonian}
\hat{H}_{F \otimes A} \equiv \: &
      \sum_{n}
      \left\{
        n \omega \ket{n}\bra{n}_F \otimes \hat{I}_A
      \right\} \nonumber \\
    & +
    \sum_{m}
      \left\{
         \hat{S}_{F}(m) \otimes \tilde{H}_A(m)
      \right\},
\end{align}
where the ``shift operators'' are defined as:
\begin{equation} \label{eq:shift_def}
\hat{S}_F(m)  \equiv \sum_n \ket{n+m} \bra{n}_F.
\end{equation}
The original time-dependent problem has now been formulated as a familiar {\em time-independent} eigenvalue problem by which the quasi-energies $E_j$'s and expansion coefficients (the $D_j(m,\alpha)$'s in Eq.~\ref{eq:quasi_components}) may be determined.

Since there are an infinite number of Fourier coefficients, the matrix representation of $\hat{H}_{F \otimes A}$ is infinite, reflecting a superfluity associated with the quasi-states and quasi-energies \cite{shortdoi:czxdv8}: if we shift a quasi-energy by $\hbar \omega$ --- or equivalently by $\omega$ in the simplified units of this section --- this may be compensated for by simultaneously shifting the corresponding expansion coefficients, so as to describe the same solution.
i.e.~we may combine Eq.'s  \ref{eq:quasi_components} and \ref{eq:time_evolving_qs} to give:
\begin{equation} \label{eq:shift_properties}
\ket{\psi_j(t)}_A = e^{-i E_{j,p} t} \sum_{m,\alpha}
  e^{i m \omega t} D_j( m - p, \alpha) \ket{\alpha}_A
\end{equation}
where $E_{j,p} \equiv E_{j} + p\omega$ with $p$ {\em any} integer. (By convention we may choose $-\omega/2 < E_{j} \le \omega/2$ for all $j$.)  The corresponding shifted eigenvectors of $\hat{H}_{F \otimes A}$ are given by
\begin{equation}
\ket{\phi_{j,p}}_{F \otimes A} \equiv
  \left[
  \left(
    \sum_{\ell} \ket{\ell+p} \bra{\ell}_F
  \right)
  \otimes
  \hat{I}_A
  \right]
  \ket{\phi_j}_{F \otimes A}.
\end{equation}
Examination of $\hat{H}_{F \otimes A}$ shows that if $E_j$ and $\ket{\phi_{j}}$ are an eigenpair then so are $E_{j,p}$ and $\ket{\phi_{j,p}}$.

Thus, although matrix representations of $\hat{H}_{F \otimes A}$ are infinite (due to the $F$ space), there are really only $N_A$ non-trivially distinct eigenpairs, which is consistent with the finite summation of Eq.~\ref{eq:floquet_operator}. In practice, estimates of the spectrum of $\hat{H}_{F \otimes A}$ may be obtained through diagonalization in a truncated, finite basis, as will be illustrated by an example in Section \ref{se:sft_example}.

\vspace{10pt} 
\subsubsection{The SFT propagator}
\label{se:sft_propagator}

Shirley \cite{shortdoi:czxdv8} showed that it is possible to express the matrix elements of the unitary time evolution operator using the Floquet Hamiltonian $\hat{H}_{F \otimes A}$ directly, without explicit reference to the quasi-energies and states:
\begin{widetext}
\begin{equation} \label{eq:shirley_propagator}
\braket{\beta|\hat{U}_A(t,0)|\alpha}_A =
  \sum_n e^{i n \omega t}
    \left[ \bra{n}_F \otimes \bra{\beta}_A
    \right]
    e^{-i \hat{H}_{F \otimes A} t}
    \left[ \ket{0}_F \otimes \ket{\alpha}_A
    \right].
\end{equation}
\end{widetext}
Although $\alpha$ and $\beta$ represent arbitrary atomic states, in a slight abuse of terminology we will refer this expression as a {\em propagator}.  It follows from the insertion of the form for $\ket{\phi_j(t)}_A$ given by Eq.~\ref{eq:general_fa_solution} into the expression for the unitary time evolution operator given by Eq.~\ref{eq:floquet_operator} (see Supplemental Materials \cite{supplemental}).  Together with the definition of the SFT Hamiltonian (Eq.~\ref{eq:floquet_hamiltonian}) it encapsulates all of SFT, and thus will serve as a useful means by which to compare SFT and MMFT.

\subsubsection{Example of the usage of SFT}
\label{se:sft_example}

To illustrate our main points regarding the correctness of MMFT we will consider computation of the time evolution of an atomic system with a Hamiltonian consisting of two periodic, commensurate couplings.  In this section we look at a specific example using SFT, and will return to the same example using MMFT in Sections \ref{se:mmft_example} and \ref{se:mmft_ts_example}.  Our particular choice of system is simple and subfield-agnostic, but otherwise is somewhat arbitrary. (Although we are ultimately interested in bichromatic microwave dressing of Rydberg atoms \cite{shortdoi:gc82gc}, that is not relevant here.  And although we choose frequencies such that $N_1=1$ and $N_2=2$, other choices, such as
$N_1=2$ and $N_2=3$, would also illustrate our points.)

The atomic system is described using an orthonormal basis consisting of two states, lower ($\ell$) and upper ($u$), evolving according to the TDSE (Eq.~\ref{eq:tdse}) with the Hamiltonian:\footnote{The unperturbed energy level splitting $E_{u}-E_{\ell}$ results in ``equal and opposite'' detunings of $\omega$ and $2\omega$, and is inspired by Ref.~\cite{wos:1979GH00700003}, but is of no special significance to our main points.
}
\begin{align}
\hat{H}_A(t) & = E_u \ket{u}\bra{u}_A + E_{\ell} \ket{\ell}\bra{\ell}_A
\tag*{} \\ \\& \quad
\begin{aligned}[b]
+ & 2 V \left(
       \cos ( \omega t) + \cos(2\omega t + \phi_{2\omega})
     \right) \\
& \times
     (\ket{u} \bra{l}_A + \ket{l} \bra{u}_A)
\end{aligned} \label{eq:example_hamiltonian}
\end{align}
where $E_u = 3/2$, $E_{\ell}=0$, $\omega=1$, $V=1$ and $\hbar=1$. We will study this $\omega,2\omega$ system with different values of the phase $\phi_{2\omega}$, as it turns out to be significant in the comparison of SFT and MMFT.

\begin{figure}[!htbp]
\includegraphics{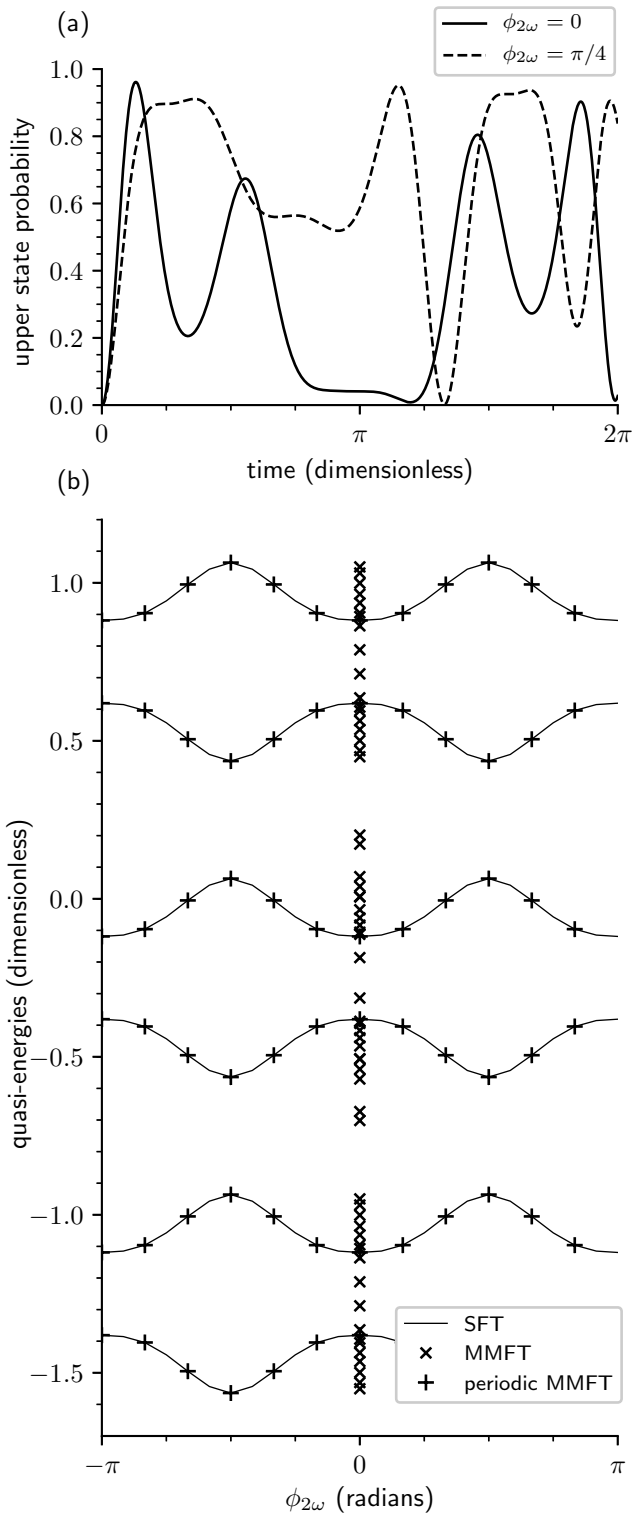}
\caption{\label{fg:sft_example}(a) Time evolution of the upper state population given the Hamiltonian of Eq.~\ref{eq:example_hamiltonian}, with all population initially in the ground state.
(b) Partial spectra for the same physical system as (a), computed by diagonalization of the SFT Hamiltonian with varying $\phi_{2\omega}$ ($\boldsymbol{-}$ lines), computed by diagonalization of the MMFT Hamiltonian for $\phi_{2\omega}=0$ ($\times$ points) as described in Section \ref{se:mmft_example}, and computed by diagonalization of the MMFT Hamiltonian using periodic boundary conditions ($+$ points) as described in Section \ref{se:mmft_ts_example}.
}
\end{figure}

With such a small atomic space ($N_A=2$) it is straightforward to directly integrate the TDSE with the Hamiltonian of Eq.~\ref{eq:example_hamiltonian}, using standard numerical methods, without any consideration of Floquet theory.  Starting with all the population in $\ell$ at $t=0$, Fig.~\ref{fg:sft_example}a) illustrates the computed time evolution for two values of the phase $\phi_{2 \omega}$.

This time evolution may also be computed using the SFT propagator of Eq.~\ref{eq:shirley_propagator}, where the plotted quantity in Fig.~\ref{fg:sft_example}a) is $\left|\bra{u} \hat{U}_A(t) \ket{\ell}_A \right|^2$.  For $\hat{H}_{F \otimes A}$ we use Eq.~\ref{eq:floquet_hamiltonian}, with
\begin{subequations}
\begin{align}
\tilde{H}_A(0) &= E_u \ket{u}\bra{u}_A + E_{\ell} \ket{\ell}\bra{\ell}_A \\
\tilde{H}_A(\pm 1) &= V (\ket{u} \bra{l}_A + \ket{l} \bra{u}_A) \\
\tilde{H}_A(\pm 2) &= V e^{\pm i \phi_{2\omega}}
  (\ket{u} \bra{l}_A + \ket{l} \bra{u}_A),
\end{align}
\end{subequations}
and all other couplings zero.

The $F \otimes A$ space of SFT is infinite-dimensional due to the Fourier decomposition space $F$.  To numerically evaluate Eq.~\ref{eq:shirley_propagator} we truncate the standard basis for $F$.
Instead of summation over all integer $n$, only a finite set is considered:
 $\mathcal{N} = \{ n \in \mathbb{Z} \mid n_{\text{min}} \le n \le n_{\text{max}} \}$:
the basis for $F \otimes A$ being formed from the tensor product of the vectors for $F$ from $\mathcal{B}_F = \{ \ket{n}_F \mid n \in \mathcal{N} \}$ and the basis vectors for the atomic space.  The size of the basis is $N_{\mathcal{B}_F} \times N_A$ where $N_{\mathcal{B}_F}$ refers to the number of elements in $\mathcal{B}_F$.   A finite matrix version of $\hat{H}_{F \otimes A}$ is considered by simply ignoring couplings between vectors not described by this finite basis.  This finite-dimensional version of $\hat{H}_{F \otimes A}$ is diagonalized numerically and in place of $e^{-i \hat{H}_{F \otimes A} t}$ in Eq.~\ref{eq:shirley_propagator}, we use
$\sum_j  e^{-i E_j t} \ket{\phi_j}\bra{\phi_j}_{F \otimes A}$
where $j$ indices a complete set of eigenpairs of the finite $\hat{H}_{F \otimes A}$.

If for simplicity\footnote{This straightforward approach is not the most efficient means to numerically compute unitary time evolution using SFT.  A more judicious choice of $\mathcal{B}_F$ and exploitation of the ``repeated'' nature of the spectrum would improve efficiency.} we select $\mathcal{B}_F$'s with $n_{\text{min}} = - n_{\text{max}}$, then $n_{\text{max}} \ge 10$ is necessary for the finite matrix version of Eq.~\ref{eq:shirley_propagator} to compute $\left|\bra{u} \hat{U}_A(t) \ket{\ell}_A \right|^2$ at $t=2\pi$ to within $10^{-2}$ for $\phi_{2\omega}=0$.  Under these conditions, the results of direct integration of the TDSE and the computation using SFT are visually indistinguishable over the full time interval $0$ to $2\pi$ in Fig.~\ref{fg:sft_example}a).

Figure \ref{fg:sft_example}b) shows a finite portion of the computed quasi-energy spectrum (the eigenvalues of the finite $\hat{H}_{F \otimes A}$).  As expected based on the discussion around Eq.~\ref{eq:shift_properties} the quasi-energies repeat vertically in the figure with a periodicity of $\hbar \omega$ ($=1$ for the simplified units of this example).  (This property is approximate with a finite basis for $F$.)

Based on the significant difference in the time evolution observed in Fig.~\ref{fg:sft_example}a) for the two values of $\phi_{2\omega}$ we might expect that the quasi-energy spectrum depends on $\phi_{2\omega}$.   This is confirmed in part b) of the figure, where the quasi-energy spectrum is plotted as a function of $\phi_{2\omega}$ (by repeatedly diagonalizing the finite $\hat{H}_{F \otimes A}$ as $\phi_{2\omega}$ is varied).

This example may also be treated using MMFT, as discussed in the next section.

\subsection{Many-mode Floquet theory (MMFT)}    \label{se:mmft}

\subsubsection{The MMFT Hamiltonian and propagator}

For concreteness and correspondence with a common experimental scenario, consider an atomic system with dipole coupling to the electric field.  With the superposition of two sinusoidal fields:
\begin{equation} \label{eq:two_electric_coupling_example}
\hat{H}_{A}(t) = \tilde{H}_A(0)
  - \vec{\mu}_A \cdot \vec{E}_1 \cos(\omega_1 t + \phi_1)
  - \vec{\mu}_A \cdot \vec{E}_2 \cos(\omega_2 t + \phi_2).
\end{equation}
As Leasure \cite{shortdoi:b7ng8m} pointed out and we have discussed in the introduction, if $\omega_1/\omega_2$ may be expressed as the ratio of two integers $N_1/N_2$ then such a Hamiltonian has a single periodicity.\footnote{In all that follows we will assume that $N_1$ and $N_2$ are positive integers with a greatest common divisor of one.} With the common ``base frequency'' $\omega_B = \omega_1 / N_1 = \omega_2 / N_2$, the two time-dependent couplings in Eq.~\ref{eq:two_electric_coupling_example} are simply couplings at different harmonics of $\omega_B$, so that we may Fourier expand:
\begin{equation}
\hat{H}_A(t) = \sum_{m} \tilde{H}_A(m) e^{i m \omega_B t}
\end{equation}
and thus the entire approach of Shirley \cite{shortdoi:czxdv8} is applicable.
(The example of Section \ref{se:sft_example} corresponds to $N_1=1$, $N_2=2$.)

Ho {\it et al.}~\cite{shortdoi:fg723q} take this idea as their starting point for MMFT, and then consider ``relabelling'' Fourier basis vectors in Shirley's formulation as basis vectors from the tensor product of {\em two} Fourier spaces:
\begin{equation} \label{eq:relabel}
\ket{n}_F \xrightarrow{\text{relabel}} \ket{n_1}_{F_1} \otimes \ket{n_2}_{F_2},
\end{equation}
where $n \omega_B = n_1 \omega_1 + n_2 \omega_2$; or equivalently $n=n_1 N_1 + n_2 N_2$.  We will discuss shortly whether or not this relabelling is possible for all $n$, and if so, if the choice of $n_1$ and $n_2$ is unique.  In any case the new basis to be used consists of all possible integer $n_1$ and $n_2$'s (in principle;  in practice the basis is truncated using convergence criteria).

The paper introducing MMFT \cite{shortdoi:fg723q} focused on time-dependent Hamiltonians in the form of Eq.~\ref{eq:two_electric_coupling_example}.  Since then, the MMFT terminology has come to refer to a slightly more general situation in which the time-dependent Hamiltonians of interest have the form:
\begin{equation} \label{eq:general_h_fourier_expansion}
\hat{H}_A(t) = \sum_{p,q} \tilde{H}_A(p,q) e^{i(p \omega_1 + q \omega_2)t}.
\end{equation}
for which Eq.~\ref{eq:two_electric_coupling_example} may be considered a special case (see the example of Section \ref{se:mmft_example}
).  We will focus on the two-mode\footnote{Depending on the context we will refer to the modes as {\em frequencies}, {\em fields}, or {\em couplings}, having in mind typical Hamiltonians of the form of Eq.~\ref{eq:two_electric_coupling_example}.  Arguably a more precise terminology for MMFT is {\em many-frequency Shirley Floquet theory}.} case for concreteness (see, for example, Ref.~\cite{shortdoi:dw9t7w} for a many-mode generalization).

In this more general version of MMFT, the time-independent MMFT Hamiltonian in the new $F_1 \otimes F_2 \otimes A$ space is given as (with $\hbar =1$):
\begin{align} \label{eq:general_mmft_hamiltonian}
\hat{H}_{F_1 \otimes F_2 \otimes A} &=
  \sum_{n_1, n_2}
    \left[
       n_1 \omega_1 + n_2 \omega_2
    \right]
    \ket{n_1}\bra{n_1}_{F_1} \nonumber \\
    & \qquad \quad \otimes
    \ket{n_2}\bra{n_2}_{F_2}
    \otimes
    \hat{I}_A \nonumber \\
    & \quad +
    \sum_{p,q} \hat{S}_{F_1}(p) \otimes \hat{S}_{F_2}(q)
       \otimes \tilde{H}_A(p, q)
\end{align}
with the $\hat{S}_{F_1}$ and $\hat{S}_{F_2}$ shift operators defined by Eq.~\ref{eq:shift_def}.

Ho {\it et al.}~\cite{shortdoi:fg723q} generalize (but do not prove) the propagator due to Shirley \cite{shortdoi:czxdv8} (our Eq.~\ref{eq:shirley_propagator}) to:
\begin{align} \label{eq:mmft_propagator}
\bra{\beta}
\hat{U}(t,0)
\ket{\alpha}
 = \: &
 \sum_{n_1, n_2} \begin{aligned}[t] &
 e^{i \left[n_1 \omega_1 + n_2 \omega_2 \right] t} \nonumber \\
 & \left[
   \bra{n_1}_{F_1} \otimes \bra{n_2}_{F_2} \otimes \bra{\beta}_A
 \right] \\
 & e^{-i \hat{H}_{F_1 \otimes F_2 \otimes A} t} \\
 & \left[
 \ket{0}_{F_1} \otimes \ket{0}_{F_2} \otimes \ket{\alpha}_A \right].
  \end{aligned} \\
\end{align}
This expression appears again in the literature following Ho {\it et al.}~\cite{shortdoi:fg723q}, in, for example, Ref.'s \cite{shortdoi:gfth9n,shortdoi:cfhh}.  Both this propagator and the form of the MMFT Hamiltonian appear to be plausible generalizations of the analogous well-established results of SFT (Eq.'s \ref{eq:floquet_hamiltonian} and \ref{eq:shirley_propagator}).  Furthermore, the MMFT Hamiltonian has the desirable property that no explicit references to $N_1$ and $N_2$ appear, so that its structure remains unchanged if $\omega_1$ and $\omega_2$ are varied.
But we are not aware of a prior resolution of the issues that we discuss in the next section.

\subsubsection{Concerns with the validity of MMFT}
\label{se:concerns}

As mentioned in the introduction, concerns have been raised regarding the validity of MMFT \cite{shortdoi:dcs7gd,shortdoi:cmtcwh}.  One troubling aspect of Ho et al.'s \cite{shortdoi:fg723q} justification for MMFT is the ``relabelling'' process (Eq.~\ref{eq:relabel}).  Specifically,  given any integer $n$, are there always integers $n_1$ and $n_2$ satisfying $n_1 N_1 + n_2 N_2 = n$, and if so, is the solution unique?  Ho {\it et al.}~\cite{shortdoi:fg723q} discuss existence but not uniqueness.  Here we note that for a given rational $\omega_1/\omega_2$, the corresponding $N_1$ and $N_2$ can always be chosen so that their greatest common divisor $\operatorname{gcd}(N_1,N_2)$ is one.  Thus there is {\em always} a solution (see, for example, Ref.~\cite{rosen_note}).\footnote{
As such,  $\operatorname{gcd}(N_1,N_2)=1$ implies that only the $p=0$ blocks of Ho {\it et al.}~\cite{shortdoi:fg723q} are necessary (see the discussion following their Eq.~10). For this reason, we do not make use of their ``$p$-block'' construction.  A related discussion appears in Ref.~\cite{shortdoi:cfhh}.
}
Moreover, there are an {\em infinite} number of solutions i.e.~given one solution for integers $n_1$ and $n_2$ satisfying: $n = n_1 N_1 + n_2 N_2$,
we also have:
\begin{equation} \label{eq:multiple_solutions}
\underbrace{(n_1 + \ell N_2)}_{n'_1} N_1 +
\underbrace{(n_2 - \ell N_1)}_{n'_2} N_2 = n
\end{equation}
for all integer $\ell$, giving an infinite number of solutions ($n'_1$ and $n'_2$) (and also all possible solutions).    Thus the relabelling process is not unique --- basis states of different $n_1$ and $n_2$ can correspond to the same $n$, raising the question of over-completeness of the standard $n_1,n_2$ MMFT basis \cite{shortdoi:dcs7gd, shortdoi:cfhh}.  We are not able to see any straightforward way to address this specific deficiency in Ho et al.'s \cite{shortdoi:fg723q} derivation, which has been characterized as incomplete \cite{shortdoi:cfhh}.

A related issue is that for Hamiltonians like Eq.~\ref{eq:two_electric_coupling_example} it has been pointed out that the eigenvalues of the MMFT Hamiltonian do not depend on the relative phase of the two fields \cite{shortdoi:cmtcwh} (we detail this argument later in Section \ref{se:significance_of_k_ne_0}).  Our example in Section \ref{se:sft_example} and Fig.~\ref{fg:sft_example} shows that this independence is problematic, as the quasi-energies obtained from SFT clearly {\em do} depend on $\phi_{2\omega}$.

Although Ho {\it et al.}~\cite{shortdoi:fg723q} provided a specific numerical example showing that MMFT reproduces the results of explicit time integration of the TDSE, the effective $N_1$ and $N_2$'s were quite large (when considered in conjunction with coupling strengths).  In this situation previous workers have described MMFT as being {\em approximately} correct (see, for example, Ref.~\cite{shortdoi:dcs7gd}), as typical finite basis sets used would not contain any ``repeated states''.

However these favourable conditions are not present in our example $\omega,2\omega$ system of Section \ref{se:sft_example}.  Surprisingly, the next section empirically illustrates that MMFT works.

\subsubsection{Example of the usage of MMFT}

\label{se:mmft_example}

The MMFT propagator can be numerically evaluated in an analogous manner as the SFT propagator, as was described in Section \ref{se:sft_example}.  The difference being that we need to truncate the basis for $F_1 \otimes F_2$, rather than for $F$.  Thus in Eq.~\ref{eq:mmft_propagator} we will take the summations of a finite set of $n_1$ and $n_2$'s.  Similarly, a finite version of $\hat{H}_{F_1 \otimes F_2 \otimes A}$ can be diagonalized numerically to evaluate the matrix elements of $e^{-i \hat{H}_{F_1 \otimes F_2 \otimes A} t}$.

\begin{figure*}
\includegraphics[width=6in]{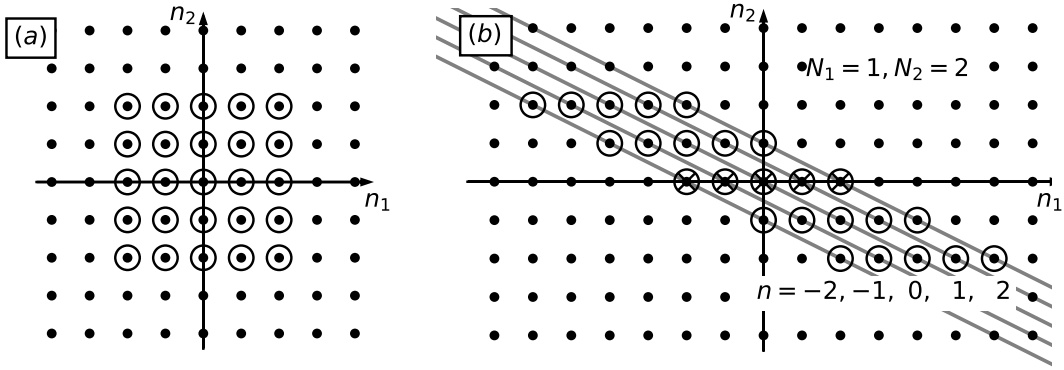}
\caption{\label{fg:truncated_basis_sets_for_f1_f2_illustration}
Finite basis sets for the $F_1 \otimes F_2$ space used in MMFT calculations.  Points on the integer lattice ($\cdot$) represent basis vectors $\ket{n_1}_{F_1} \otimes \ket{n_2}_{F_2}$.  In principle, basis sets for MMFT calculations should run over all integer $n_1$ and $n_2$; however finite basis sets --- consisting of basis vectors marked by ($\odot$) --- are typically used to numerically diagonalize MMFT Hamiltonians. Shown are: (a) a conventional choice (e.g.~Ref.~\cite{shortdoi:fg723q}), and
(b) a basis set suitable for maintaining the ``translational invariance'' of the MMFT Hamiltonian (Eq.~\ref{eq:mmft_hamiltonian_translational_invariance}).  The basis vector selection in (b) depends on $N_1$ and $N_2$ ($N_1=1$ and $N_2=2$ in this case).  The lines connect basis vectors corresponding to the same $n$.  The canonical vectors $\ket{n_1(n)}_{F_1} \otimes \ket{n_2(n)}_{F_2}$ for each $n$ (see Appendix \ref{se:canonical_choice}) are indicated ($\otimes$).
}
\end{figure*}

The time evolution of the $\omega,2\omega$ system of Section \ref{se:sft_example} may also be determined using MMFT, with $\omega_1=1$, $\omega_2=2$,
\begin{subequations}
\begin{align}
\tilde{H}_A(0,0) &= E_u \ket{u}\bra{u}_A + E_{\ell} \ket{\ell}\bra{\ell}_A \\
\tilde{H}_A(\pm 1,0) &= V (\ket{u} \bra{l}_A + \ket{l} \bra{u}_A) \\
\tilde{H}_A(0,\pm 1) &= V e^{\pm i \phi_{2\omega}}
  (\ket{u} \bra{l}_A + \ket{l} \bra{u}_A)
\end{align}
\end{subequations}
and all other couplings zero.  We construct a finite basis for $F_1 \otimes F_2$ with basis kets of the form $\ket{n_1}_{F_1} \otimes \ket{n_2}_{F_2}$ for all $n_1$ and $n_2$ such that $n_1 \in \mathcal{N}$ and $n_2 \in \mathcal{N}$, with $\mathcal{N} = \{ n \in \mathbb{Z} \mid -n_{\text{max}} \le n \le n_{\text{max}} \}$.  See Fig.~\ref{fg:truncated_basis_sets_for_f1_f2_illustration}(a) for an example of a finite basis set with $n_{\text{max}}=2$.

We find that $n_{\text{max}} \ge 9$ is necessary for the finite matrix version of Eq.~\ref{eq:mmft_propagator} to compute $\left|\bra{u} \hat{U}_A(t) \ket{\ell}_A \right|^2$ at $t=2\pi$ to within $10^{-2}$ for $\phi_{2\omega}=0$. Under these conditions, the results of direct integration of the TDSE and the computation using MMFT are visually indistinguishable over the full time interval $0$ to $2\pi$ in Fig.~\ref{fg:sft_example}a).

That MMFT may accurately compute the time evolution in this system was a surprise to us given the concerns of the previous section and the nature of the eigenvalues of the finite basis MMFT Hamiltonian.  Specifically, Fig.~\ref{fg:sft_example}b) shows the eigenvalues for the truncated MMFT Hamiltonian with $\phi_{2\omega}=0$ (the $\times$ points distributed vertically at $\phi_{2\omega}=0$) illustrating that the spectrum of the MMFT Hamiltonian {\em does not} correspond to the SFT quasi-energies (solid line) at $\phi_{2 \omega}=0$.  Despite this discrepancy, in numerical experimentation on a variety of commensurate systems (e.g.~$2\omega, 3\omega$) we have found that Eq.~\ref{eq:mmft_propagator} may be used to compute unitary time evolution.

\section{The relationship of MMFT to SFT}
\label{se:equivalence}

\subsection{Equivalence of the MMFT and SFT propagators}

We will now show why calculations using the MMFT propagator given in Eq.~\ref{eq:mmft_propagator} with the MMFT Hamiltonian of Eq.~\ref{eq:general_mmft_hamiltonian} are correct for commensurate frequencies, despite the concerns discussed in Section \ref{se:concerns} and the discrepancy between the SFT and MMFT spectra noted in the previous section.  We avoid the problematic relabelling procedure of Ho {\it et al.}~\cite{shortdoi:fg723q} and
take a rather different approach.

Specifically, we will exploit a symmetry of the MMFT Hamiltonian to help show the correctness of the MMFT propagator.  Consider a unitary operator, that produces a ``translated'' version of a vector $\ket{n_1}_{F_1} \otimes \ket{n_2}_{F_2}$ corresponding to the same $n$ ($\equiv n_1 N_1 + n_2 N_2$):
\begin{equation} \label{eq:t_f1_f2_def}
\hat{T}_{F_1 \otimes F_2} \equiv \hat{S}_{F_1}(N_2) \otimes \hat{S}_{F_2}(-N_1).
\end{equation}
where the $\hat{S}$ operators are of the same form as Eq.~\ref{eq:shift_def}.  Defining $T_{F_1 \otimes F_2 \otimes A} \equiv \hat{T}_{F_1 \otimes F_2} \otimes \hat{I}_A$, it may be verified that the MMFT Hamiltonian given by Eq.~\ref{eq:general_mmft_hamiltonian} is invariant under this translation:
\begin{equation} \label{eq:mmft_hamiltonian_translational_invariance}
T_{F_1 \otimes F_2 \otimes A}^{-1}
\hat{H}_{F_1 \otimes F_2 \otimes A}
T_{F_1 \otimes F_2 \otimes A}
= \hat{H}_{F_1 \otimes F_2 \otimes A}.
\end{equation} \newline
This symmetry suggests an analogy with the tight-binding Hamiltonians used for solid-state crystals, in which every lattice site has equivalent couplings to its neighbours (see, for example, Ref.~\cite{isbn:978-0-03-083993-1}).  In the case of MMFT with commensurate frequencies, the implications of this symmetry do not appear to have been fully explored (see, for example, the pedagogical treatment of MMFT in Ref.~\cite{shortdoi:cjk7r4}).\footnote{Both Ref.'s \cite{shortdoi:gcgdhz} and \cite{shortdoi:gfwhfg} consider this analogy, but with quite different and more sophisticated objectives, focusing on incommensurate frequencies and topological aspects.}

In particular, since $\hat{T}_{F_1 \otimes F_2 \otimes A}$ commutes with $\hat{H}_{F_1 \otimes F_2 \otimes A}$, if $\ket{\psi}_{F_1 \otimes F_2 \otimes A}$ is an eigenvector of $\hat{T}_{F_1 \otimes F_2 \otimes A}$ with eigenvalue $e^{-ik}$, with $k$ real, then $\hat{H}_{F_1 \otimes F_2 \otimes A} \ket{\psi}_{F_1 \otimes F_2 \otimes A}$ is also an eigenvector of $\hat{T}_{F_1 \otimes F_2 \otimes A}$ with the same eigenvalue --- the MMFT Hamiltonian does not ``connect'' eigenvectors of $\hat{T}_{F_1 \otimes F_2 \otimes A}$ corresponding to different eigenvalues.  This suggests that we partially diagonalize $\hat{H}_{F_1 \otimes F_2 \otimes A}$ by replacing the $n_1, n_2$ basis for the $F_1 \otimes F_2$ space with one in which $\hat{T}_{F_1 \otimes F_2}$ is diagonal.  We will refer to this new basis for the $F_1 \otimes F_2$ space as the $n,k$ basis.

The $n,k$ basis vectors may be understood as the superposition of vectors of different $n_1$ and $n_2$, but the same $n$ ($\equiv n_1 N_1 + n_2 N_2$) forming eigenvectors of $\hat{T}_{F_1 \otimes F_2}$ (with eigenvalues $e^{-ik}$):
\begin{equation} \label{eq:k_basis_def}
\ket{n,k}_{F_1 \otimes F_2} =
  \frac{1}{\sqrt{N}} \sum_{p}
  e^{ipk} \: \hat{T}^p_{F_1 \otimes F_2} \ket{n_1(n)}_{F_1} \otimes \ket{n_2(n)}_{F_2}
\end{equation}
where for each $n$ we define a {\em canonical} vector: $\ket{n_1(n)}_{F_1} \otimes \ket{n_2(n)}_{F_2}$ satisfying $n=n_1(n) N_1 + n_2(n) N_2$.  One approach to making a specific choice for $n_1(n)$ and $n_2(n)$ is given in Appendix \ref{se:canonical_choice}.  The summation may be considered as a limit taken as $N$, the number of terms in the summation over $p$, goes to infinity.  We do not belabor taking this limit, as it may be avoided, as shown in Appendix \ref{se:non_truncated_justification}.  Imagining the summation as finite is helpful for obtaining an intuitive understanding of the MMFT and SFT equivalence.  Furthermore, in Section \ref{se:mmft_ts_example} we will show that satisfactory numerical implementations of MMFT can be obtained using finite summations over $p$ while preserving the symmetry of the MMFT Hamiltonian given by Eq.~\ref{eq:mmft_hamiltonian_translational_invariance}.

In the $n,k$ basis, the final bras in the MMFT propagator of Eq.~\ref{eq:mmft_propagator} correspond to $k=0$:
\begin{align}
& \sum_{n_1,n_2} e^{i(n_1 \omega_1 + n_2 \omega_2) t }
\bra{n_1}_{F_1} \otimes \bra{n_2}_{F_2}
\nonumber \\
& =
\sqrt{N} \: \sum_n
 e^{i \omega_B n t} \bra{n,k=0}_{F_1 \otimes F_2}.
\end{align}
The $F_1 \otimes F_2$ part of the initial ket may be written as a superposition of different $k$ vectors:\footnote{Again we are making use of the convenient fiction that $N$ is finite --- in principle the superposition of Eq.~\ref{eq:this_should_be_an_integral} should be expressed as an integral with $k$ ranging continuously from $-\pi$ to $\pi$.}
\begin{equation} \label{eq:this_should_be_an_integral}
\ket{0}_{F_1} \otimes \ket{0}_{F_2}
 = \frac{1}{\sqrt{N}}
   \sum_k \ket{n=0,k}_{F_1 \otimes F_2}.
\end{equation}
But since $\hat{H}_{F_1 \otimes F_2 \otimes A}$ does not couple vectors of different $k$, the final bras dictate that only the $k=0$ term in the initial ket superposition is relevant, allowing us to write:
\begin{widetext}
\begin{equation} \label{eq:propagator_with_k_equal_zero}
\bra{\beta}
\hat{U}_A(t)
\ket{\alpha}_A
 =
 \sum_{n}
 e^{i n \omega_B t}
 \left[
   \bra{n,k=0}_{F_1 \otimes F_2} \otimes \bra{\beta}_A
 \right]
 e^{-i \hat{H}_{F_1 \otimes F_2 \otimes A} t}
 \left[
 \ket{n=0,k=0}_{F_1 \otimes F_2} \otimes \ket{\alpha}_A \right].
\end{equation}
\end{widetext}
Thus we see that the part of the spectrum of $\hat{H}_{F_1 \otimes F_2 \otimes A}$
corresponding to $k \ne 0$ is irrelevant to the propagator.  In essence this is the origin of the controversy over MMFT:  although the spectrum of $\hat{H}_{F_1 \otimes F_2 \otimes A}$ contains the appropriate Floquet quasi-energies and states ($k=0$), it also contains extraneous eigenpairs (corresponding to $k \ne 0$).  However, the propagator ``selects'' the relevant eigenvectors i.e.~those corresponding to $k=0$.

To complete the argument for the correctness of Eq.~\ref{eq:mmft_propagator}, it must be shown that Eq.~\ref{eq:propagator_with_k_equal_zero} --- which is the same as \ref{eq:mmft_propagator} but rewritten using the $n,k$ basis --- reproduces Shirley's propagator, Eq.~\ref{eq:shirley_propagator}.  For this purpose it is sufficient to show that for all possible atomic states specified by $\gamma$ and $\nu$, and all integer $n'$ and $n''$'s, the following equality between matrix elements holds:
\begin{widetext}
\begin{equation} \label{eq:required_matrix_element_equality}
\left( \bra{n',k=0}_{F_1 \otimes F_2} \otimes \bra{\gamma}_A
\right)
  \hat{H}_{F_1 \otimes F_2 \otimes A}
\left(
\ket{n'',k=0}_{F_1 \otimes F_2} \otimes \ket{\nu}_A
\right)
=
\bra{n'}_F \otimes \bra{\gamma}_A
  \hat{H}_{F \otimes A} \ket{n''}_F \otimes \ket{\nu}_A .
\end{equation}
\end{widetext}
The preceding equality follows from rewriting the $k=0$ bra and ket of the LHS in the $n_1,n_2$ basis using Eq.~\ref{eq:k_basis_def} and then substituting $\hat{H}_{F_1 \otimes F_2 \otimes A}$ from Eq.~\ref{eq:general_mmft_hamiltonian}.  We also use
\begin{equation} \label{eq:h_sft_mmft_correspondence}
\tilde{H}_A(r) = \sum_{p,q} \delta_{r,pN_1 + qN_2} \tilde{H}_A(p,q).
\end{equation}
within $\hat{H}_{F \otimes A}$ (from Eq.~\ref{eq:floquet_hamiltonian})
on the RHS of Eq.~\ref{eq:required_matrix_element_equality}.

As the correctness of SFT is well-established, and we have just shown that for commensurate frequencies the MMFT and SFT propagators are equivalent (see also Appendix \ref{se:non_truncated_justification}), we conclude that usage of the MMFT propagator (Eq.~\ref{eq:mmft_propagator}) is correct for commensurate frequencies.

\subsection{The significance of the $k \ne 0$ eigenvectors of the MMFT Hamiltonian} \label{se:significance_of_k_ne_0}

Now let us address an objection to the use of MMFT for commensurate frequencies raised by Potvliege and Smith \cite{shortdoi:cmtcwh}, who pointed out that a change in the relative phase of two commensurate fields can be written as a unitary transformation of the MMFT Hamiltonian, and thus its eigenvalues are independent of relative phase (shown below).

This independence seems at odds with experimental observations that the behavior of quantum systems in the presence of external perturbing fields of $\omega$ and $2 \omega$ depends on the relative phase of the two fields (see, for example, Ref.~\cite{shortdoi:b4s5gq} and the references in Ref.~\cite{isbn:978-3-527-40904-4}).  Our $\omega,2\omega$ example certainly exhibits this dependence (Fig.~\ref{fg:sft_example}): the time evolution depends strongly on $\phi_{2\omega}$, as do the quasi-energies computed using SFT.

We resolve this apparent paradox by observing that the unitary transformation corresponding to changing the relative phase of the fields is essentially a translation in ``$k$-space'', so that a different portion of the spectrum of $\hat{H}_{F_1 \otimes F_2 \otimes A}$ is ``moved'' into $k=0$ (recall that the propagator only makes use of the $k=0$ part of the spectrum).  Diagonalization of $\hat{H}_{F_1 \otimes F_2 \otimes A}$ may be viewed as a computation of the quasi-energy spectra for {\em all} phases of the two fields.  (In a finite basis this is only approximately realized --- a numerical example will be provided in Section \ref{se:mmft_ts_example}.)

To justify the preceding claim, let us consider time-dependent Hamiltonians written in terms of two phases $\phi_1$ and $\phi_2$:
\begin{equation} \label{eq:phases_general_h_fourier_expansion}
\hat{H}_A(t) = \sum_{p,q} \tilde{H}_A(p,q) e^{ip(\omega_1 t + \phi_1)+iq(\omega_2 t + \phi_2)},
\end{equation}
which incorporates Eq.~\ref{eq:two_electric_coupling_example} as a special case.  The corresponding MMFT Hamiltonian is:
\begin{align} \label{eq:phases_general_mmft_hamiltonian}
& \hat{H}_{F_1 \otimes F_2 \otimes A} (\phi_1, \phi_2) = \nonumber \\
& \:\:\:  \sum_{n_1, n_2}
    \left[
       n_1 \omega_1 + n_2 \omega_2
    \right]
    \ket{n_1}\bra{n_1}_{F_1}
    \otimes
    \ket{n_2}\bra{n_2}_{F_2}
    \otimes
    \hat{I}_A \nonumber \\
    & +
    \sum_{p,q}
    e^{i(p \phi_1 + q \phi_2)}
    \hat{S}_{F_1}(p) \otimes \hat{S}_{F_2}(q)
       \otimes \tilde{H}_A(p, q),
\end{align}
where we have explicitly indicated the phase-dependence for comparison with the original MMFT Hamiltonian with no phase shifts: $\hat{H}_{F_1 \otimes F_2 \otimes A}(0,0)$.

Defining
\begin{equation}
\hat{U}_F(\phi) \equiv \sum_n e^{-in\phi} \ket{n}\bra{n}_F
\end{equation}
we may make use of the identity:
$e^{i p \phi} \hat{S}_F(p) = \hat{U}_F(\phi)^{-1} \hat{S}_F(p) \hat{U}_F(\phi)$
for $F_1$ and $F_2$ in the last term of Eq.~\ref{eq:phases_general_mmft_hamiltonian} to write:
\begin{align}
\hat{H}_{F_1 \otimes F_2 \otimes A}(\phi_1, \phi_2) = &
\left[
  \hat{U}_{F_1}(\phi_1)^{-1} \otimes \hat{U}_{F_2}(\phi_2)^{-1}
    \otimes \hat{I}_A
\right] \nonumber \\
&
\hat{H}_{F_1 \otimes F_2 \otimes A}(0, 0) \nonumber \\
&
\left[
  \hat{U}_{F_1}(\phi_1) \otimes \hat{U}_{F_2}(\phi_2)
    \otimes \hat{I}_A
\right],
\end{align}
justifying the claim \cite{shortdoi:cmtcwh} that a change in the phases of the fields corresponds to a unitary transformation of the MMFT Hamiltonian.  As a consequence, given an eigenvector $\ket{\psi}_{F_1 \otimes F_2 \otimes A}$ of $\hat{H}_{F_1 \otimes F_2 \otimes A}(0,0)$, it is also true that  $\hat{U}_{F_1}(\phi_1)^{-1} \otimes \hat{U}_{F_2}(\phi_2)^{-1} \otimes \hat{I}_A \ket{\psi}_{F_1 \otimes F_2 \otimes A}$ is an eigenvector of $\hat{H}_{F_1 \otimes F_2 \otimes A}(\phi_1, \phi_2)$ with the same eigenvalue.

Using the $n,k$ basis vectors given by Eq.~\ref{eq:k_basis_def}, together with the convention of Appendix \ref{se:canonical_choice}, we may determine how $\hat{U}_{F_1}(\phi_1)^{-1} \otimes \hat{U}_{F_2}(\phi_2)^{-1}$ effects a shift in $k$-space:
\begin{align}
& \hat{U}_{F_1}(\phi_1)^{-1} \otimes \hat{U}_{F_2}(\phi_2)^{-1}
\ket{n,k}_{F_1 \otimes F_2}
  \nonumber \\
& = e^{i(n_1(n) \phi_1 + n_2(n) \phi_2)}
\ket{n,k+N_2 \phi_1 - N_1 \phi_2}_{F_1 \otimes F_2} \nonumber \\
& =
e^{i n (n_1(1) \phi_1 + n_2(1) \phi_2)}
\ket{n,k+N_2 \phi_1 - N_1 \phi_2}_{F_1 \otimes F_2}.
\end{align}
Thus, the quasi-energies for non-zero $\phi_1$ and $\phi_2$ are the eigenvalues of $\hat{H}_{F_1 \otimes F_2 \otimes A}(0, 0)$ corresponding to\footnote{The case of $\phi_1 \ne 0$ and $\phi_2 \ne 0$ but yet $N_2\phi_1 - N_1\phi_2=0$ corresponds to an identical time-translation for both fields --- the quasi-energies are unchanged and the quasi-states are time-shifted.  Equation \ref{eq:k_phase_relationship} defines what we mean by {\em relative} phase.}
\begin{equation} \label{eq:k_phase_relationship}
k = N_1\phi_2 - N_2 \phi_1,
\end{equation}
as these $k \ne 0$ eigenvalues of $\hat{H}_{F_1 \otimes F_2 \otimes A}(0, 0)$ correspond to the $k = 0$ eigenvalues of $\hat{H}_{F_1 \otimes F_2 \otimes A}(\phi_1, \phi_2)$.
We will show an example of this correspondence in Section \ref{se:mmft_ts_example}.

Pivotal to the preceding argument has been the point that not all eigenvalues of the MMFT Hamiltonian correspond to quasi-energies (for a fixed set of field phases).  Thus the suggestion \cite{shortdoi:dcs7gd} that for commensurate frequencies the eigenvalues of the MMFT Hamiltonian represent ``phase-averaged'' quasi-energies is not generally correct.  Of course if the eigenvalues are phase-independent then they will be phase averages.  The analogous situation for the tight-binding Hamiltonian is that at high-interatomic spacings/low-overlap the energies simply become the atomic energies --- different $k$'s are energy degenerate.  For MMFT with commensurate frequencies, large $N_1$ and $N_2$ and weak couplings will have a similar effect.

\subsection{Example of the usage of MMFT with retention of translational symmetry (periodic boundary conditions)} \label{se:mmft_ts_example}

Although the $F_1 \otimes F_2$ space used to write two-mode MMFT Hamiltonians is infinite, the example of Section \ref{se:mmft_example} illustrated that satisfactory numerical solutions for time evolution may be obtained using a truncated basis set for this space --- provided it is sufficiently large.  However, in a truncated basis set the MMFT Hamiltonian will not typically exhibit the translational symmetry of Eq.~\ref{eq:mmft_hamiltonian_translational_invariance} {\em exactly}.  As such, $k$ may no longer be considered to be a good quantum number of the quasi-states computed by diagonalization of this Hamiltonian.

In this section we show that a judicious selection of a finite set of $n_1,n_2$ basis vectors, together with the application of periodic boundary conditions --- analogous to those used in models of solid-state crystals --- preserves the translational symmetry of the MMFT Hamiltonian {\em exactly} in a finite $n_1,n_2$ basis.  Transforming from this basis to one in which $k$ is a good quantum number block diagonalizes the MMFT Hamiltonian and allows us to illustrate the connection between the $k \ne 0$ eigenpairs and the relative phase of the fields, as discussed in the previous section.

Recall that each $n_1,n_2$ basis vector has a single associated $n$ ($\equiv n_1 N_1 + n_2 N_2$), but that for a given $n$ there are an infinite number of associated $n_1,n_2$ vectors (see the discussion surrounding Eq.~\ref{eq:multiple_solutions}).  Selection of an appropriate finite basis amounts to deciding which $n$'s will be represented in the basis, and then choosing a finite number of $n_1,n_2$ vectors for each of these $n$'s (Fig.~\ref{fg:truncated_basis_sets_for_f1_f2_illustration}(b) provides an example).  More specifically, an algorithm for the selection of a finite basis set for $F_1 \otimes F_2$ is:
\begin{enumerate}
\item choose a finite set of integers $\mathcal{N}$ specifying the $n$'s that will be represented by the basis.   This will typically be the same set as would be used for an equivalent SFT calculation (see Section \ref{se:sft_example}).  For example, $\mathcal{N} = \{ n \in \mathbb{Z} \mid n_{\text{min}} \le n \le n_{\text{max}} \}$  and in Fig.~\ref{fg:truncated_basis_sets_for_f1_f2_illustration}(b), $\mathcal{N}= \{ -2,-1,0,1,2 \}$, corresponding to each diagonal line.
\item for each $n \in \mathcal{N}$ decide on a {\em canonical} $n_1,n_2$ basis vector, denoted as $\ket{n_1(n)}_{F_1} \otimes \ket{n_2(n)}_{F_2}$.  One way to make this choice is given in Appendix \ref{se:canonical_choice} and an example is shown in Fig.~\ref{fg:truncated_basis_sets_for_f1_f2_illustration}(b) (using $\otimes$ markers).
\item for each $n$, generate a set of $n_1,n_2$ basis vectors by repeated application of $\hat{T}_{F_1 \otimes F_2}$ (see Eq.~\ref{eq:t_f1_f2_def}) and/or its inverse (both of which preserve $n$) to the canonical basis vector for this $n$, giving the basis set:
$\mathcal{B}_{F_1 \otimes F_2} = \{ \hat{T}_{F_1 \otimes F_2}^{\ell} \ket{n_1(n)}_{F_1} \otimes \ket{n_2(n)}_{F_2} \mid n \in \mathcal{N} \land \ell \in \mathcal{L} \}$
where $\mathcal{L} \equiv \{ \ell \in \mathbb{Z} \mid \ell_{\text{min}} \le \ell \le \ell_{\text{max}} \}$.
In Fig.~\ref{fg:truncated_basis_sets_for_f1_f2_illustration}(b), $\mathcal{L}= \{ -2,-1,0,1,2 \}$, with each element corresponding to a different location along the diagonals.
\end{enumerate}
The finite basis $\mathcal{B}_{F_1 \otimes F_2}$ generated by the preceding procedure has the following property:  given {\em any} $n_1, n_2$ basis vector with corresponding $n \in \mathcal{N}$, there always exists one unique integer $q$ such that
$(\hat{T}_{F_1 \otimes F_2}^{N_{\mathcal{L}}})^{q} \ket{n_1}_{F_1} \otimes \ket{n_2}_{F_2}$ is an element of $\mathcal{B}_{F_1 \otimes F_2}$, where $N_{\mathcal{L}}$ is the number of elements in the set $\mathcal{L}$ (if the $n_1,n_2$ vector is already contained within $\mathcal{B}_{F_1 \otimes F_2}$ then $q=0$).  Each vector within $\mathcal{B}_{F_1 \otimes F_2}$ may be considered as defining an equivalence class containing elements that are not within $\mathcal{B}_{F_1 \otimes F_2}$ (in addition to the vector within $\mathcal{B}_{F_1 \otimes F_2}$).

These equivalences allow periodic boundary conditions to be implemented:  if a term in the MMFT Hamiltonian couples a vector $n_1, n_2$ from $\mathcal{B}_{F_1 \otimes F_2}$ to $n_1', n_2'$, and this vector $n_1', n_2'$ may be ``translated'' --- as described in the previous paragraph --- to $n_1'',n_2''$ within $\mathcal{B}_{F_1 \otimes F_2}$ (always possible if $n_1'N_1 + n_2'N_2 \in \mathcal{N}$), then this coupling is counted as a contribution towards the matrix element between $n_1, n_2$ and $n_1'', n_2''$; otherwise it is ignored.  Stated in another way: we implement periodic boundary conditions by taking matrix elements of $(\hat{C}_{F_1 \otimes F_2} \otimes I_A) \hat{H}_{F_1 \otimes F_2 \otimes A}$ and $(\hat{C}_{F_1 \otimes F_2} \otimes I_A) \hat{T}_{F_1 \otimes F_2 \otimes A}$ where $\hat{C}_{F_1 \otimes F_2} \equiv \sum_{q \in \mathbb{Z}} (\hat{T}_{F_1 \otimes F_2}^{N_{\mathcal{L}}})^{q}$.  When the finite matrix representations are constructed in this manner, they exhibit the symmetry of Eq.~\ref{eq:mmft_hamiltonian_translational_invariance}.  In the rest of this section we will refer to
$T_{F_1 \otimes F_2 \otimes A}$, $T_{F_1 \otimes F_2}$, and $H_{F_1 \otimes F_2 \otimes A}$ (note no hats) as the finite matrix versions of their operator counterparts with periodic boundary conditions applied.

After $H_{F_1 \otimes F_2 \otimes A}$ has been written in the finite basis formed by combining $\mathcal{B}_{F_1 \otimes F_2}$ with the atomic basis, we may rewrite it in a new basis in which $k$ is a good quantum number.  Since $H_{F_1 \otimes F_2 \otimes A}$ does not connect basis vectors of differing $k$, the Hamiltonian will be block diagonal in this new basis --- with each block and its eigenpairs corresponding to a specific $k$. The new basis may be derived from $\mathcal{B}_{F_1 \otimes F_2}$ using Eq.~\ref{eq:k_basis_def}, which we can make precise by specifying that the summation is over all $p \in \mathcal{L}$, $N$ is replaced by $N_{\mathcal{L}}$, and $\hat{T}_{F_1 \otimes F_2}$ is replaced by its periodic version.  Equation \ref{eq:k_basis_def} then takes the form of a discrete Fourier transform and $T_{F_1 \otimes F_2}$ has eigenvalues uniformly spaced around the unit circle in the complex plane.  Following convention, these eigenvalues may be written as $e^{-ik}$ with $k=2\pi j / N_{\mathcal{L}}$ where $j$ an integer ranging from $-N_{\mathcal{L}}/2$ to $N_{\mathcal{L}}/2-1$ if $N_{\mathcal{L}}$ is even, or $-(N_{\mathcal{L}}-1)/2$ to $(N_{\mathcal{L}}-1)/2$ if $N_{\mathcal{L}}$ is odd. 
We have implemented the preceding procedure for the $\omega, 2\omega$ example discussed in Sections \ref{se:sft_example} and \ref{se:mmft_example}.  In Fig.~\ref{fg:sft_example}b), the $+$ points represent the eigenvalues of $H_{F_1 \otimes F_2 \otimes A}$, where we have used the correspondence $\phi_{2\omega}=k$ from Eq.~\ref{eq:k_phase_relationship} with $N_1=1$, $N_2=2$ and $\phi_1=0$ suitable for the Hamiltonian of Eq.~\ref{eq:example_hamiltonian}.  The finite basis used for $F_1 \otimes F_2$ has $N_{\mathcal{L}} = 12$ and $\mathcal{N} = \{-8,-7, \ldots, 8 \}$.  
Recall that the solid lines of Fig.~\ref{fg:sft_example}(b) correspond to SFT computations with varying $\phi_{2\omega}$ (where the SFT Hamiltonian is constructed and diagonalized for each $\phi_{2\omega}$).  By comparison with the $+$ points, we see that diagonalization of a {\em single} MMFT Hamiltonian samples quasi-energies for a discrete set of relative phases.  The spectrum calls to mind the analogy with solid-state crystals: as $N_{\mathcal{L}} \rightarrow \infty$ the spectrum of the MMFT Hamiltonian ceases to have isolated eigenvalues, but rather becomes band-like (this property has been previously noted by Potvliege and Smith \cite{shortdoi:cmtcwh}). 
We do not advocate use of the procedure of this section (a special basis set and periodic boundary conditions) for any practical computations, as each $k$ block of the MMFT Hamiltonian is essentially an SFT Hamiltonian corresponding to a certain relative phase.  Our purpose in this section was to illustrate with a specific example the connection between the $k$ labelling of eigenpairs of the MMFT Hamiltonian and the phases of the fields.

\section{Summary and discussion}    \label{se:discussion}

For commensurate frequencies, the MMFT Hamiltonian has a ``translational'' symmetry (Eq.~\ref{eq:mmft_hamiltonian_translational_invariance}) analogous to that found in tight binding models of solid-state crystals.  Using this symmetry we have established that when applied to time-dependent periodic Hamiltonians involving two commensurate frequencies (of the form given by Eq.~\ref{eq:general_h_fourier_expansion}):\footnote{Although we have focused on the two-mode case for concreteness, similar conclusions apply to MMFT in cases of more than two modes.}
\begin{enumerate}[(1)]
\item the MMFT propagator for unitary time evolution (Eq.~\ref{eq:mmft_propagator}) as originally given by Ho {\it et al.}~\cite{shortdoi:fg723q} using the MMFT Hamiltonian (in the modern form of Eq.~\ref{eq:general_mmft_hamiltonian}) is correct, but
\item not all of the eigenpairs of the MMFT Hamiltonian correspond to the Floquet quasi-energies and quasi-states, and
\item ``invalid'' eigenpairs of the MMFT Hamiltonian correspond to the quasi-energies and quasi-states for different time-dependent Hamiltonians.  These different Hamiltonians correspond to those arising from relative phase shifts of the fields contributing to the Hamiltonian (as detailed in Section \ref{se:significance_of_k_ne_0} and illustrated by the example of the $\omega,2\omega$ system in Section \ref{se:mmft_ts_example}).
\end{enumerate}
Although point (1) appears to be a confirmation of Ref.~\cite{shortdoi:fg723q}, one of the authors of Ref.~\cite{shortdoi:fg723q} --- following Ref.'s \cite{shortdoi:dcs7gd} and \cite{shortdoi:cmtcwh} --- later restricted the application of MMFT to incommensurate frequencies, treating the commensurate case using SFT \cite{shortdoi:b6r5dc} (as we have done in Section \ref{se:sft_example} for the $\omega,2\omega$ example).  It appears that authors who reference the original MMFT paper are not always aware of this restriction (partially erroneous because of point (1) and partially correct because of point (2)) and the concerns with the validity of MMFT that have been raised in the literature \cite{shortdoi:dcs7gd,shortdoi:cmtcwh,shortdoi:cfhh}.

Point (2) is important since it is normal (and correct) to take the eigenvalues and eigenstates of Shirley's SFT Hamiltonian (Eq.~\ref{eq:floquet_hamiltonian}) as corresponding to the Floquet quasi-energies and quasi-states, whereas this is not necessarily correct for MMFT.  Although one must be slightly cautious when diagonalizing the SFT Hamiltonian within a finite basis, the problematic eigenpairs appear at the extremes of the spectrum.  By contrast, as the $\omega,2\omega$ example of Fig.~\ref{fg:sft_example}b) shows (the X points), erroneous --- as they do not correspond to the Floquet quasi-states --- eigenpairs of the MMFT Hamiltonian can appear in the centre of the spectrum.  Some (in the ``bands'') correspond (approximately) to different phases of the fields, whereas others (those in the ``gaps'') are artifacts of basis set truncation.

That some MMFT eigenpairs correspond to the quasi-energies for different relative phases of the fields may be an interesting observation (point (3)), but not necessarily useful.  In a finite basis, extra eigenpairs corresponding to differing phases of the fields imply a larger matrix representation of the MMFT Hamiltonian than necessary.  If one emulates the translational symmetry of the MMFT Hamiltonian (Eq.~\ref{eq:mmft_hamiltonian_translational_invariance}) in a finite basis using periodic boundary conditions to allow block diagonalization (as we have done for illustrative purposes in Section \ref{se:mmft_ts_example}), the result is simply equivalent to application of SFT repeatedly for a discrete set of relative phases.

Just as a tight binding Hamiltonian with negligible couplings between lattice sites will produce a set of degenerate atomic energies (the bands collapsing to isolated energies), it is also the case that depending on $N_1$ and $N_2$ and the couplings, the approximate diagonalization of MMFT Hamiltonians using finite basis sets may give the correct quasi-energies.   In fact, we have not been able to find any examples in the literature where MMFT has given incorrect quasi-energies --- presumably because those studies, like the original MMFT paper \cite{shortdoi:fg723q}, have concentrated on large $N_1$ and $N_2$'s, and weak couplings.  We are not yet aware of how to state these criteria precisely.

Finally, let us consider MMFT and our results from a modern perspective.  Two periodic ``dressing'' fields can be used to engineer a quantum system, optimizing properties such as low sensitivity to decohering fields \cite{shortdoi:gc82gc}.  For numerical optimization, the MMFT Hamiltonian has the seemingly(!)~attractive property that its structure does not explicitly depend on the precise ratio of the two field frequencies.  By contrast, Shirley's formalism is more cumbersome, as the SFT Hamiltonian structure depends on the exact rational representation of the frequency ratio (i.e.~$N_1$ and $N_2$).  If the dressing frequencies are to be varied as part of an optimization process, then the simplicity of MMFT is appealing, but ultimately problematic --- optimization may lead to frequency ratios corresponding to low $N_1$ and $N_2$.  In this context, our $\omega,2\omega$ example sounds a warning: naive interpretation of the MMFT Hamiltonian eigenenergies as quasi-energies may be incorrect.\footnote{To apply SFT to commensurate multiple frequency problems, the choice of efficient basis sets may still be inspired by MMFT: select {\em some} of the harmonics of the base frequency using $n=n_1 N_1 + n_2 N_2$, where $n_1$ and $n_2$ are small integers, checking for and eliminating(!)~any repeated $n$'s.}  This warning is despite the correctness of the MMFT propagator (Eq.~\ref{eq:mmft_propagator}) using the same Hamiltonian.

\begin{acknowledgments}
We thank A.~Cooper-Roy and N.~Fladd for comments on this manuscript.  This work was supported by the Natural Sciences and Engineering Research Council of Canada.
\end{acknowledgments}

\begingroup

\appendix

\section{Choice of canonical $n_1$, $n_2$}
\label{se:canonical_choice}

In the main text, we have referred at points (e.g.~Eq.~\ref{eq:k_basis_def}) to vectors: $\ket{n_1(n)}_{F_1} \otimes \ket{n_2(n)}_{F_2}$, specific to each $n$, satisfying $n_1(n) N_1 + n_2(n) N_2 = n$.  Here we describe a method to select these vectors. i.e.~how to choose $n_1$ and $n_2$ for a given $n$ (the functional dependence on $N_1$ and $N_2$ is left implicit in our notation).

The extended Euclidean algorithm (EEA) (see for example Ref.~\cite{rosen_note}) simultaneously determines both the greatest common divisor (gcd) of two positive integers $a$ and $b$ and a specific integer solution for $x$ and $y$ satisfying $ax + by = \operatorname{gcd}(a,b)$.  Since for any given rational frequency ratio we may always choose $N_1$ and $N_2$ so that $\operatorname{gcd}(N_1, N_2) = 1$, we use the EEA to solve for $n_1(1)$ and $n_2(1)$ satisfying
\begin{equation} \label{eq:n_one}
n_1(1) N_1+n_2(1) N_2 = 1
\end{equation}
(and also verify that $\operatorname{gcd}(N_1,N_2)=1$).
Multiplying both sides of Eq.~\ref{eq:n_one} by $n$ suggests that we define:
$n_1(n) \equiv n_1(1) n$ and $n_2(n) \equiv n_2(1) n$. This choice is used in Fig.~\ref{fg:truncated_basis_sets_for_f1_f2_illustration}(b) and in the numerical example of Section \ref{se:mmft_ts_example}.

Reference \cite{shortdoi:gf3wqr} points out that the EEA produces an integer solution for $x$ and $y$ to $ax+by=\operatorname{gcd}(a,b)$ having minimal $x^2 + y^2$, which is desirable for the aesthetics of Fig.~\ref{fg:truncated_basis_sets_for_f1_f2_illustration}(b), but by no means necessary.

\vspace{10pt} 

\section{Justification of the MMFT propagator without basis set truncation}  \label{se:non_truncated_justification}

In the main text, the equivalence of the MMFT propagator (Eq.~\ref{eq:mmft_propagator}) to Shirley's Floquet propagator (Eq.~\ref{eq:shirley_propagator}) for commensurate frequencies is demonstrated using physically suggestive summations over a finite number of $n_1,n_2$ basis vectors to produce $n,k$ vectors.  Here we justify the equivalence of the propagators in a more rigorous manner.

The MMFT propagator (Eq.~\ref{eq:mmft_propagator}) can be written in a form resembling the SFT propagator through the introduction of two linear maps: 1) a ``promotion'' map $P$ from $F \otimes A$ to $F_1 \otimes F_2 \otimes A$, and 2) a ``demotion'' map $D$ from $F_1 \otimes F_2 \otimes A$ to $F \otimes A$:
\begin{align} \label{eq:mmft_propagator_promote_demote}
\bra{\beta}
\hat{U}_A(t)
\ket{\alpha}
 =
 \sum_{n}
 e^{i n \omega_B t}
&
   \bra{n}_{F} \otimes \bra{\beta}_A
 \:
 D
 \:
 e^{-i \hat{H}_{F_1 \otimes F_2 \otimes A} t}
 \:
 P
 \nonumber \\
&
 \ket{0}_{F} \otimes \ket{\alpha}_A,
\end{align}
with
\begin{equation}
D \equiv \sum_{n_1,n_2} \ket{n_1 N_1 + n_2 N_2}_F
      \bra{n_1}_{F_1} \otimes \bra{n_2}_{F_2}
    \otimes \hat{I}_A,
\end{equation}
and
\begin{equation}
P \equiv \sum_n \ket{n_1(n)}_{F_1} \otimes \ket{n_2(n)}_{F_2}
       \bra{n}_F \otimes \hat{I}_A,
\end{equation}
where $n_1(n) N_1 + n_2(n) N_2 = n$ (see Appendix \ref{se:canonical_choice}; the choice of $P$ is not unique and nor is it required to be).  Note that although
\begin{equation} \label{eq:dp_identity}
D P = \hat{I}_{F \otimes A},
\end{equation}
we have:
\begin{equation}
P D \ne \hat{I}_{F_1 \otimes F_2 \otimes A},
\end{equation}
since mapping from $F_1 \otimes F_2 \rightarrow F$ ``loses'' information i.e.~it is possible that $D \ket{n_1}_{F_1} \otimes \ket{n_2}_{F_2} = D \ket{n'_1}_{F_1} \otimes \ket{n'_2}_{F_2}$ with $n_1 \ne n'_1$ or $n_2 \ne n'_2$.  Applying $P$ to map back into $F_1 \otimes F_2$ does not restore this information.

Comparison of the MMFT propagator written using $D$ and $P$ (Eq.~\ref{eq:mmft_propagator_promote_demote}) to the SFT propagator (Eq.~\ref{eq:shirley_propagator}) shows that their equivalence will follow if
\begin{equation} \label{eq:powers_of_h_equivalence}
\hat{H}_{F \otimes A}^j = D \hat{H}_{F_1 \otimes F_2 \otimes A}^j P
\end{equation}
for all non-negative integer $j$.  The $j=0$ case follows from Eq.~\ref{eq:dp_identity}.  For $j>0$ it is sufficient that
\begin{equation} \label{eq:hd_dh}
\hat{H}_{F \otimes A} D = D \hat{H}_{F_1 \otimes F_2 \otimes A},
\end{equation}
since by acting with $P$ from the right on both sides (and using Eq.~\ref{eq:dp_identity}) we have
\begin{equation}
\hat{H}_{F \otimes A} = D \hat{H}_{F_1 \otimes F_2 \otimes A} P,
\end{equation}
and subsequently acting from the left of both sides with $H_{F \otimes A}$ and using \ref{eq:hd_dh} to simplify the RHS gives Eq.~\ref{eq:powers_of_h_equivalence} for $j=2$.  This process may be continued to establish Eq.~\ref{eq:powers_of_h_equivalence} for any positive integer $j$.

To show Eq.~\ref{eq:hd_dh} we take $\hat{H}_{F_1 \otimes F_1 \otimes A}$ from Eq.~\ref{eq:general_mmft_hamiltonian}, and  $\hat{H}_{F \otimes A}$ as given by Eq.~\ref{eq:floquet_hamiltonian}, making use of Eq.~\ref{eq:h_sft_mmft_correspondence} to ensure that both SFT and MMFT Hamiltonians refer to the same time-dependent Hamiltonian in the atomic space.
This establishes the equivalence of the MMFT propagator (Eq.~\ref{eq:mmft_propagator}) to Shirley's Floquet propagator (Eq.~\ref{eq:shirley_propagator}).

\endgroup

\bibliography{bib_files/bibtex_from_zotero,bib_files/supplemental,bib_files/rosen_note}

\IfSubStringInString{\detokenize{generated}}{\jobname}{
\verbatiminput{git_info_generated.txt}
}{}

\end{document}


\pagebreak
\widetext
\clearpage
\begin{center}
\textbf{\large Supplemental Materials: Many-mode Floquet theory with commensurate frequencies}
\end{center}
\setcounter{equation}{0}
\setcounter{figure}{0}
\setcounter{table}{0}
\setcounter{page}{1}
\setcounter{section}{0}
\makeatletter
\renewcommand{\theequation}{S\arabic{equation}}
\renewcommand{\thefigure}{S\arabic{figure}}
\renewcommand{\bibnumfmt}[1]{[S#1]}
\renewcommand{\citenumfont}[1]{S#1}

\section{Floquet theory for unitary time evolution}

\label{se:supp_floquet_derivation}

Here we show  --- making no claim of originality --- how Eq.~\ref{eq:floquet_operator} of the main text (a special case of Floquet theory) may be considered to arise {\em constructively} from computation of the unitary time evolution operator in a way that makes use of the periodicity of the Hamiltonian.  See also Ref.~\cite{shortdoi:gfw8sp}. \\

Time evolution from $t'=0$ to $t'=t$ may be decomposed into two stages:
\begin{enumerate}
\item evolution from $t'=0$ to $t' = n T$, where $n \equiv \lfloor t/T \rfloor$, where $\lfloor t /T \rfloor$ is considered to be the largest integer less than or equal to $t/T$ (following the notation of Ref.~\cite{isbn:978-0-201-55802-9}), followed by
\item evolution from $t' = n T$ to $t'=t$, corresponding to a duration of $t \operatorname{mod} T \equiv t-nT$ (again following Ref.~\cite{isbn:978-0-201-55802-9}).
\end{enumerate}
With this decomposition:
\begin{equation} \label{eq:u_decompose}
\hat{U}(t,0) = \hat{U}(t, n T) \:\:
\hat{U}(n T,0).
\end{equation}

Due to the periodic nature of the Hamiltonian (invariance under a shift in time by $T$), we have $\hat{U}(t_2,t_1) = \hat{U}(t_2+NT,t_1+NT)$ for all $t_1$ and $t_2$, and any integer $N$, so that $\hat{U}(t, n T) = \hat{U}(t- n T, 0) = \hat{U}(t \operatorname{mod} T, 0)$ and also
\begin{align} \label{eq:steps}
  \hat{U}(n T,0) &= \hat{U}(n T, (n-1) T) \: \ldots \: \hat{U}(2T, T) \:\: \hat{U}(T, 0) \\
             &= \hat{U}(T,0)^{n}
\end{align}
allowing us to write Eq.~\ref{eq:u_decompose} as
\begin{equation} \label{eq:powers_of_u}
\hat{U}(t,0) = \hat{U}(t \operatorname{mod} T, 0) \:\: \hat{U}(T,0)^{n}.
\end{equation}

With $\hat{H}_A$ and $\hat{U}$ operators on a finite-dimensional inner-product space, we now consider their finite $N_A \times N_A$ matrix representations in an orthonormal basis  --- dropping the $\hat{}$ (hat) symbol --- and use two standard results of matrix theory (see, for example, Ref.~\cite{isbn:978-1-138-19989-7}): 1) any unitary matrix $U$ may be written as $U=\exp(-i M)$ where $M$ is Hermitian, and 2) any Hermitian matrix $M$ may be written as $WDW^{\dagger}$ where $W$ is a unitary matrix and $D$ is a real diagonal matrix.
Using these results, we are permitted to write $U(T,0) = \exp(-i WDW^{\dagger})$, or equivalently --- since $W$ is unitary --- $U(T,0) = W \exp(-iD) W^{\dagger}$.

The time evolution operator of Eq.~\ref{eq:powers_of_u} may now be represented as:
\begin{align}
  U(t,0)
  & = U(t \operatorname{mod} T, 0)
  \left[W e^{-iD} W^{\dagger}
    \right]^{n}.
\end{align}
Recalling that $n$ is an integer and that $W^{\dagger} W = I$, we have
\begin{align}
  U(t,0) & = U(t \operatorname{mod} T, 0) W e^{-iDn} W^{\dagger}.
\end{align}
Since $t = n T  + (t \operatorname{mod} T)$ by the definition of both $n \equiv \lfloor t/T \rfloor$ and $t \operatorname{mod} T$, we may write $n = (t - (t \operatorname{mod} T))/T$, and thus
\begin{align}
  U(t,0) & =
  U(t \operatorname{mod} T, 0) W
  e^{-iD \left[ (t - (t \operatorname{mod} T))/T \right]}
  W^{\dagger} \\
  & =
  U(t \operatorname{mod} T, 0) W
  e^{iD (t \operatorname{mod} T)/T}
  e^{-iD t / T}
  W^{\dagger}.
\end{align}
Defining the unitary matrix:
\begin{equation} \label{eq:def_phi}
  \Phi(t) \equiv U(t \operatorname{mod} T, 0) W
  e^{iD (t \operatorname{mod} T)/T}
\end{equation}
and noting that $(t + T) \operatorname{mod} T = t \operatorname{mod} T$, we have
\begin{equation} \label{eq:floquet}
U(t,0)  = \Phi(t) e^{-iD t / T} \Phi^{\dagger}(0)
\end{equation}
where $\Phi(t) = \Phi(t+T)$ for all $t$.  Equation \ref{eq:floquet} is the matrix representation of Eq.~\ref{eq:floquet_operator} of the main text --- a special case of Floquet theory describing quantum-mechanical unitary time evolution due to a time-periodic and Hermitian Hamiltonian.

Each of the columns of $\Phi(t)$ may be used to construct solution vectors
corresponding to the quasi-states and quasi-energies of Eq.~\ref{eq:each_floquet_soln} of the main text:
\begin{equation} \label{eq:column_solutions}
\psi_j(t) = \Phi_{:,j}(t) e^{-i E_j t / \hbar},
\end{equation}
where $E_j = D_{j,j} \hbar / T$.

\section{Derivation of the SFT Hamiltonian}

\label{se:supp_details_of_sft_hamiltonian}

Substitution of
\begin{equation}
\ket{\psi_j(t)}_A = e^{-i E_j t}
 \sum_{n} e^{i n t} \:\:
   \left[ \bra{n}_F \otimes \hat{I}_A
   \: \right] \:\:
   \ket{\phi_j}_{F \otimes A}
\end{equation}
into the TDSE, yields:
\begin{equation}
\sum_{m} \tilde{H}_A(m) e^{i m t}
\sum_{n}
e^{i n t}
   \left[ \bra{n}_F \otimes \hat{I}_A
   \: \right] \:\:
   \ket{\phi_j}_{F \otimes A}
= \sum_{n} e^{i n t} (E_j - n)
   \left[ \bra{n}_F \otimes \hat{I}_A
   \: \right] \:\:
   \ket{\phi_j}_{F \otimes A}.
\end{equation}
Equivalently:
\begin{equation}
\sum_{p} e^{i p t}
\sum_{q} \tilde{H}_A(q)
   \left[ \bra{p-q}_F \otimes \hat{I}_A
   \: \right] \:\:
   \ket{\phi_j}_{F \otimes A}
= \sum_{n} e^{i n t} (E_j - n)
   \left[ \bra{n}_F \otimes \hat{I}_A
   \: \right] \:\:
   \ket{\phi_j}_{F \otimes A}
,
\end{equation}
which by the uniqueness property of Fourier decomposition implies that for all integer $r$:
\begin{equation}
\sum_{q} \tilde{H}_A(q)
   \left[ \bra{r-q}_F \otimes \hat{I}_A
   \: \right] \:\:
   \ket{\phi_j}_{F \otimes A}
= (E_j - r)
   \left[ \bra{r}_F \otimes \hat{I}_A
   \: \right] \:\:
   \ket{\phi_j}_{F \otimes A}
.
\end{equation}
By rearrangement and insertion of a resolution of the identity in the $F$ space, we obtain:
\begin{equation}
\left[ \bra{r}_F \otimes \hat{I}_A \right]
  \left[
    \sum_{n}
      \left\{
        n \ket{n}\bra{n}_F \otimes \hat{I}_A
      \right\}
    +
    \sum_{n, m}
      \left\{
         \ket{n + m}\bra{n}_F \otimes \tilde{H}_A(m)
      \right\}
  \right]
     \ket{\phi_j}_{F \otimes A}
  = \left[ \bra{r}_F \otimes \hat{I}_A \right]
     E_j
     \ket{\phi_j}_{F \otimes A}
\end{equation}
which, since it is true for all $r$, may be interpreted as an eigenvalue problem:
\begin{equation}
\hat{H}_{F \otimes A} \ket{\phi_j}_{F \otimes A} = E_j \ket{\phi_j}_{F \otimes A}
\end{equation}
where the {\em Floquet Hamiltonian} $\hat{H}_{F \otimes A}$ is:
\begin{equation}
   \hat{H}_{F \otimes A} \equiv
      \sum_{n}
      \left\{
        n \ket{n}\bra{n}_F \otimes \hat{I}_A
      \right\}
    +
    \sum_{n, m}
      \left\{
         \ket{n + m}\bra{n}_F \otimes \tilde{H}_A(m)
      \right\}
.
\end{equation}

\section{Derivation of the SFT propagator}

\label{se:supp_details_of_sft_propagator}

To obtain the propagator due to SFT (Eq.~\ref{eq:shirley_propagator}), we may insert the form for $\ket{\phi_j(t)}_A$ given by Eq.~\ref{eq:general_fa_solution} of the main text into the expression for the unitary time evolution operator given by Eq.~\ref{eq:floquet_operator} of the main text, to obtain:
\begin{equation}
\hat{U}_A(t) = \sum_{j,p,q} e^{i p t}
  \left[ \bra{p}_F \otimes \hat{I}_A
  \right]
  e^{-i E_{j,0} t}
  \ket{\phi_{j,0}}
  \bra{\phi_{j,0}}_{F \otimes A} \:\:
  \left[ \ket{q}_F \otimes \hat{I}_A
  \right].
\end{equation}
Making use of the shift properties of the quasi-states (see the discussion surrounding Eq.~\ref{eq:shift_properties} of the main text):
\begin{equation}
\hat{U}_A(t) = \sum_{j,p,q} e^{i p t}
  \left[ \bra{p-q}_F \otimes \hat{I}_A
  \right]
  e^{-i E_{j,0} t}
  \ket{\phi_{j,-q}}
  \bra{\phi_{j,-q}}_{F \otimes A} \:\:
  \left[ \ket{0}_F \otimes \hat{I}_A
  \right].
\end{equation}
Instead of the summation over $p$ and $q$, we can consider $q$ as the outer summation, and instead of summation over $p$ we may perform the inner sum over $n \equiv p-q$, yielding:
\begin{equation}
\hat{U}_A(t) = \sum_{j,q,n} e^{i n t}
  \left[
    \bra{n}_F \otimes \hat{I}_A
  \right] \:\:
  e^{-i E_{j,-q} t} \:\:
  \ket{\phi_{j,-q}}
  \bra{\phi_{j,-q}}_{F \otimes A}  \:\:
  \left[ \ket{0}_F \otimes \hat{I}_A
  \right].
\end{equation}
Since $\ket{\phi_{j,-q}}_{F \otimes A}$ is an eigenvector of $\hat{H}_{F \otimes A}$ with eigenvalue $E_{j,-q}$ and $\sum_{j,q} \ket{\phi_{j,-q}}\bra{\phi_{j,-q}}_{F \otimes A} = \hat{I}_{F \otimes A}$ (the identity in $F \otimes A$), the propagator immediately follows:
\begin{equation}
  \braket{\beta|\hat{U}_A(t,0)|\alpha}_A =
  \sum_n e^{i n t}
    \left[ \bra{n}_F \otimes \bra{\beta}_A
    \right]
    e^{-i \hat{H}_{F \otimes A} t}
    \left[ \ket{0}_F \otimes \ket{\alpha}_A
    \right].
\end{equation}

\bibliography{bib_files/bibtex_from_zotero}

\IfSubStringInString{\detokenize{generated}}{\jobname}{
\verbatiminput{git_info_generated.txt}
}{}